\documentclass[english,preprint,showpacs,aps]{revtex4}
\usepackage[T1]{fontenc}
\usepackage[latin1]{inputenc}
\usepackage{amsmath}
\usepackage{graphicx}
\usepackage{amssymb}
\makeatletter
\usepackage{graphicx}
\usepackage{graphicx,amsfonts}
\usepackage{dcolumn}
\usepackage{bm}
\usepackage{babel}
\makeatother
\begin{document}

\title{Consistent Interactions of the 2+1 Dimensional Noncommutative Chern-Simons Field}

\author{E. A. Asano}
\affiliation{Instituto de Fisica, Universidade de S\~ao Paulo,
Caixa Postal 66318, 05315-970, S\~ao Paulo - SP, Brazil}
\email{asano, lcbrito, mgomes, petrov, ajsilva@fma.if.usp.br}
\author{L. C. T. Brito}
\author{M. Gomes}
\author{A. Yu. Petrov}
 \altaffiliation[Also at]{ Department of Theoretical Physics,
Tomsk State Pedagogical University,
Tomsk 634041, Russia
(email: petrov@tspu.edu.ru)}
\author{A. J. da Silva}
\affiliation{Instituto de Fisica, Universidade de S\~ao Paulo,
Caixa Postal 66318, 05315-970, S\~ao Paulo - SP, Brazil}

\date{\today}

\begin{abstract}
We consider 2+1 dimensional noncommutative models of scalar and
fermionic fields coupled to the Chern-Simons field. We show that,
at least up to one loop, the model containing only a fermionic field 
in the fundamental representation minimally coupled to the
Chern-Simons field is consistent in the sense that
there are no nonintegrable infrared divergences. By contrast, dangerous
infrared divergences occur if the fermion field belongs to the adjoint
representation or if the coupling of scalar matter is considered instead. 
The superfield formulation of the supersymmetric Chern-Simons
model is also analyzed and shown to be free of nonintegrable infrared
singularities and actually finite if the matter field belongs to the 
fundamental
representation
of the supergauge group. In the case of the adjoint representation
this only happens in a particular gauge. 
\end{abstract}
\pacs{11.10.Nx, 11.15-q, 11.10.Kk, 11.10.Gh}
\maketitle

\section{Introduction}

Models containing Chern-Simons (CS) fields interacting with the matter are
very important both  for the clarification of conceptual aspects as well as  for the applications of
field theory. Partially due to the recent interest in noncommutative
theories, some properties of noncommutative CS models have been
studied 
\cite{Kaminsky,Martin,Dayi,Chandra,Anacleto,Saha,
Polychronakos,Jabbari,Wu,Park,Chu,
Silva,Das,Schweda,Pani}. 
As it happens with its
commutative nonabelian counterpart, gauge invariance of the
noncommutative CS model demands the quantization of
CS coefficient \cite{Polychronakos,Jabbari,Wu,Park}.
Up to one loop, this was proven to hold for the U(1) pure gauge model
and also when minimally coupled fermions are included \cite{Chu,Silva}.
Some results indicated that the pure CS theory
is actually a free field model \cite{Das}.

One problem that still deserves studies is the possible occurrence of
nonintegrable infrared singularities associated with the
ultraviolet/infrared (UV/IR) mixing. As known, such singularities
jeopardize the perturbative series and may lead to its breakdown.
For the pure CS model the absence of linear UV/IR mixing has been
verified up to one-loop order \cite{Schweda}.  In the present work, we
will examine various couplings of the CS field to matter determining
in what circumstances they may be consistent field theories. We begin
by considering separately the models of fermionic and scalar fields
minimally coupled to the CS field. For the case of fermionic
fields transforming in accord with the fundamental representation of
the gauge group there are no dangerous (nonintegrable) infrared
singularities. However, for the same model but with the fermionic
field belonging to the adjoint representation, there are linear
nonintegrable singularities in the radiative corrections to the gauge
field two point vertex function. The situation is still more
complicated in the case of a scalar field minimally coupled to the CS
field.  Here there are infrared singularities both for the fundamental
and the adjoint representation. In the case of the fundamental
representation linearly divergent  infrared singularities come from the
contributions to the scalar field four point vertex function whereas
in the case of the adjoint representation there are additional
infrared singularities in the two point vertex function of the gauge
field. We then show that the inclusion of an adequate Yukawa coupling
may remove the divergence if both the fermionic and the scalar fields
belong to the fundamental representation. For the scalar fields in the
adjoint representation there are infrared singularities which persist
even after the inclusion of fermions. More general interactions are
needed and thus we consider the noncommutative supersymmetric CS model
(see \cite{Lee,Gates,Avdeev} for some discussion on the quantum
dynamics of the commutative supersymmetric CS model) and prove for the
matter superfield both in the fundamental and in the adjoint
representations that, up to one loop the model is free from dangerous
infrared singularities and renormalizable. However, for the matter
superfield in the adjoint representation the absence of divergences
only happens in a particular gauge.

Our work is organized as follows. In Section \ref{section2} the
noncommutative models
of scalar and fermionic fields minimally coupled to the CS
field are introduced and our graphical notation is presented.  
The possible occurrence of dangerous (quadratic or linear) infrared
divergences is investigated first in  Section \ref{section3}, when
the matter fields belong to the fundamental representation, and
then in Section \ref{section4}, when the fields are in the adjoint
representation of the gauge group. The superfield formulation of the
noncommutative CS field coupled to a supersymmetric matter
is considered in Section \ref{section5}. A general overview and
comments of our results are presented in Section \ref{section6}.    

\section{Scalar and  fermionic matter minimally coupled to the Chern-Simons
 Field} \label{section2}

In this section we shall present some results concerning the coupling
of matter to the CS field. For both cases of scalar and
fermionic
matter fields, the pure gauge
part of the noncommutative action
is given by

\begin{eqnarray}\label{1}
S_{gauge} &=& \frac{1}{2}\int d^{3}x \left\{\epsilon^{\mu \nu \lambda}
\left(A_{\mu} \ast \partial_{\nu} A_{\lambda} + \frac{2 ie}{3}
A_{\mu}\ast A_{\nu} \ast A_{\lambda}\right) -\frac{1}{2\xi }
(\partial_{\mu }A^{\mu })(\partial _{\nu }A^{\nu })\right. \nonumber\\
&&+\partial_{\mu }\bar{c}\ast \left[\partial^{\mu }c+i\left(c\ast A^{\mu }
-A^{\mu }\ast c\right)\right]
  \Bigr\},
\end{eqnarray}

\noindent
where a generic gauge fixing ($\xi$) and the corresponding Faddeev-Popov
ghost actions have been included. The matter field actions are

\begin{eqnarray}\label{2}
S_{scalar}= -\int d^{3}x [(D^{\mu}  \varphi)^{\dag}\ast (D_{\mu} \varphi) +m^2\varphi^{\dag} \varphi],
\end{eqnarray}

\noindent
for the scalar field and

\begin{equation}\label{3}
S_{fermion}=-\int d^{3}x \bar \psi\ast(  \gamma^\mu D_\mu + M)\psi,
\end{equation}

\noindent
for the fermionic field. In this action $\psi$ denotes a two-component
Dirac field and the representation for the gamma matrices is such that
$\gamma ^{\mu }\gamma ^{\nu }=g^{\mu \nu }-\epsilon ^{\mu \nu \alpha
}\gamma _{\alpha }$, where  $\epsilon ^{\mu \nu \alpha}$ is the 
completely anti-symmetric Levi-Cività symbol. Throughout this work we
shall
use the metric $g_{11}=g_{22}=-g_{00}=1$. Furthermore, to avoid possible
unitarity problems \cite{Gomis} we shall keep the noncommutativity parameter 
$\Theta_{0i}= 0$.

In the above expressions $D_\mu{\cal O}$
is the covariant derivative of the field ${\cal O}$ and it is
given by

\begin{eqnarray}
D_{\mu}{\cal O}&=&\partial_{\mu}{\cal O}-ie {\cal O}\ast A_{\mu},  \\
D_{\mu}{\cal O}&=&\partial_{\mu}{\cal O}+ie[A_{\mu},{\cal O}]_{\ast},
\end{eqnarray}

\noindent
if the field ${\cal O}$ belongs to the fundamental and to the adjoint
representation, respectively (the Moyal commutator is defined as
$[A_\mu, \,{\cal O}]_\ast \equiv A_\mu\ast {\cal O} - {\cal O} \ast
A_\mu$). Unless for the Section V in this work we will employ the Landau gauge
by taking the limit $\xi\rightarrow 0$.

\begin{figure}
\begin{center}\includegraphics{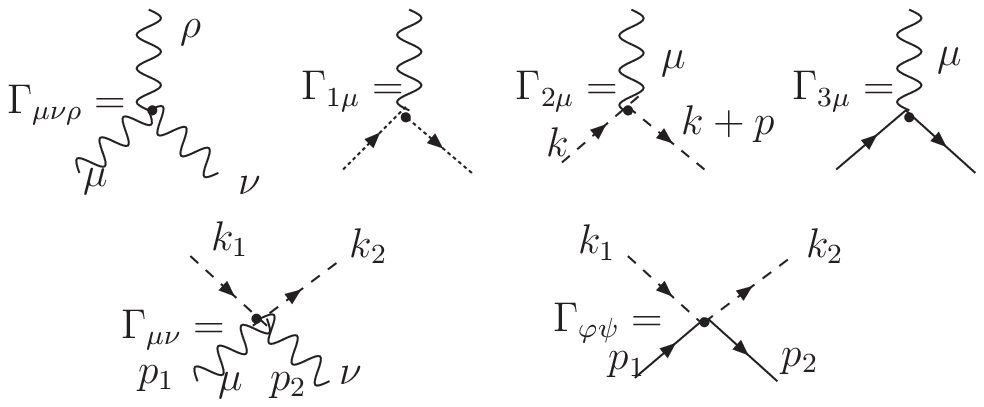}
\end{center}
\caption{Vertices for the CS field coupled to matter.
Charges flow in opposite direction to the indicated.
}
\label{Fig1}
\end{figure}

 A Feynman graph representation for the models described above
consists of wavy, continuous, dashed and dotted lines associated to
the gauge field, fermionic, scalar and ghost propagators,

\begin{eqnarray}
\Delta_{\mu\nu}(k)& =& \frac{\epsilon_{\mu\nu\rho} k^\rho}{k^2},\\
\Delta_\psi (k) &=&\frac{-i}{-i\not k +M },\\ 
\Delta_\varphi (k) &=&\frac{-i}{k^2 + m^2},\\
\Delta_c (k) &=&\frac{i}{k^2 },
\end{eqnarray}

\noindent
respectively, and of the vertices (see Fig. \ref{Fig1}):
\begin{eqnarray}
\Gamma_{\mu\nu\rho}  &=&   
2ie\,\epsilon_{\mu\nu\rho}\,\sin(k\wedge p) ,
\label{vertc}\\
\Gamma_{1\mu}& =& - 2  e k_\mu \sin(k\wedge p).
 \label{vertd}
\end{eqnarray}

The graphical correspondence for the other vertices  depends on
the representation. To distinguish the same vertex in the fundamental
and adjoint representations we include an additional index $F$ and $A$,
respectively. Thus to the trilinear scalar-gauge field vertex, indicated by
$\Gamma_{2\mu}$ in Fig.~\ref{Fig1} corresponds
\begin{equation}  
\Gamma^{F}_{2\mu} =- ie(2k+p)_\mu \,{\rm e}^{-ik\wedge p},
\label{Va}
\end{equation}

\noindent
for the fundamental representation and
\begin{equation}  
\Gamma^{A}_{2\mu} =2e(2k+p)_\mu \,\sin(k\wedge p)
\label{Va1},
\end{equation}

\noindent
for the adjoint representation. Using this convention the other
vertices are
\begin{eqnarray}
\Gamma^{F}_{\mu\nu}&=&- 2 i e^2 g_{\mu\nu} {\rm e}^{-i k_1\wedge k_2}
\cos(p_1\wedge p_2),\\
\Gamma^{A}_{\mu\nu}&=& 4 i e^2 g_{\mu\nu} \sin(k_1\wedge p_1)
\sin(k_2\wedge p_2)+ (p_1 \leftrightarrow p_2),\\
\Gamma^{F}_{3\mu}&=& -e \gamma_\mu {\rm e}^{ik\wedge p},\\
\Gamma^{A}_{3\mu}&=& 2 i e \gamma_\mu \sin(k\wedge p).
\end{eqnarray}

From these rules, the ultraviolet degree of superficial divergence
of a generic diagram $\gamma$ turns out to be
\begin{equation}
d(\gamma)= 3- N_A - N_\psi- \frac12 N_\varphi- \frac12 N_c,
\end{equation}

\noindent
where $N_A, N_\varphi $, $N_\psi$ and $N_c$ indicate the numbers of gauge,
scalar, fermionic and ghost external lines of $\gamma$ (up to one loop
$N_c=0$).

A simplifying property shared by these models  is the cancellation of
the pure gauge contributions. Thus, when computing the corrections to the
gauge field two point vertex function, one finds that  the diagrams in
Figs. \ref{Fig2}$a$ and \ref{Fig2}$b$  mutually cancel \cite{Schweda}.

Concerning the possibility of the appearance of nonintegrable infrared
singularities special care should be given to graphs with
$d(\gamma) > 0$. They can occur in the two point vertex functions of the
basic fields, in the three point vertex function $<T A_\mu
\varphi^\dagger\varphi>$ and  in the four point vertex function
$<T \varphi^\dagger\varphi^\dagger\varphi\varphi>$. In what follows we
will restrict our attention to the investigation on the possibility of
 occurrence of nonintegrable infrared singularities.

\section{Fundamental representation}
\label{section3}
Let us begin our
analysis by  considering first the case of the fundamental representation.
In this situation the one-loop contributions to the two point
functions come from  planar graphs and so do not induce infrared
nonintegrable singularities. Thus,  up to one loop the model 
whose action is $S_{gauge}+ S_{fermion}$ is renormalizable 
and  free from dangerous UV/IR mixing. 

For the scalar model described by the action $S_{gauge}+ S_{scalar}$
we need  to examine the contributions to the three and four point
vertex
functions. We have:

1. Three point vertex function. The relevant diagrams are depicted in
Fig. \ref{Fig3}. Because of properties of the Levi-Cività symbol, the
divergent parts of the integrals  associated with the graphs in the Figs.~\ref{Fig3}$a$, 
\ref{Fig3}$b$ and \ref{Fig3}$c$ actually vanish.

Furthermore, due to
our gauge choice the graphs \ref{Fig3}$d$ and \ref{Fig3}$e$ turn out to be only
logarithmically divergent and generate a mild (integrable) infrared divergence.
 
2. Four point vertex function $<T\varphi^\dagger\varphi
\varphi^\dagger\varphi>$.  There are three types of diagrams as drawn
in Figs. \ref{Fig4}$a-c$. In the Landau gauge, the diagrams in
Figs. \ref{Fig4}$a$ and \ref{Fig4}$b$ are finite but graph
\ref{Fig4}$c$ presents a linear infrared divergence as can be seen
from its analytical expression
\begin{equation}
{\rm Fig4c}= -2 e^4 {\rm e}^{i(q\wedge s+p\wedge r)} \int
  \frac{d^3k}{(2\pi)^3}\frac{\epsilon_{\mu\rho\nu} 
k^\rho}{k^2}\frac{\epsilon^{\nu\alpha\mu} 
(k+p-r)_\alpha}{(k+p-r)^2}\cos^2[k\wedge(p-r)] .
\end{equation}  

\noindent
Using $\cos^2\phi = \frac12 [1+\cos(2\phi)]$, we obtain the
following nonplanar part
\begin{eqnarray}
({\rm Fig4c})_{nplanar} &=&-2 e^4 {\rm e}^{i(q\wedge s+p\wedge r)} \int
  \frac{d^3k}{(2\pi)^3}\frac{k\cdot (k+p-r)}{k^2 (k+p-r)^2}
\cos[ 2k\wedge(p-r)]\nonumber\\
&=& \frac{i e^4}{2\pi |\tilde p -\tilde r|} + \mbox{finite term}
\end{eqnarray}

\noindent
where  in the last line $p \simeq r$. 
Of course, ``finite term'' designates the contributions that stay
finite when $p \rightarrow r$.  Although innocuous at this point the
above infrared linear divergence ruins the perturbative expansion as
it is illustrated by the graph in Fig. \ref{Fig5}, which presents a
strong nonintegrable singularity at $k=p$.  To cancel such singularity
we enlarge the model by coupling a fermionic field to the scalar field
through the following Yukawa like self-interaction
\begin{equation}
S_{Yukawa} = g \int d^3 x[\overline \psi\ast \psi\ast
  \varphi^\dagger\ast\varphi-\varphi^\dagger\ast \psi\ast\overline \psi\ast \varphi 
].\label{4}
 \end{equation}

\noindent
The relative minus sign between the terms in this expression was
chosen so that it  provides a mechanism for the cancellation of the infrared
singularity and does not vanish in the commutative limit. 
To see how this happens notice that this interaction
generates the vertex $\Gamma_{\varphi\psi}$ indicated in
Fig. \ref{Fig1},
\begin{equation} 
\Gamma_{\varphi\psi}=2 i g  \cos(k_1\wedge k_2+p_1\wedge p_2).
\end{equation}

Among the new diagrams produced by this new interaction we have the
graph in Fig. 4$d$ which gives the nonplanar contribution

\begin{equation}
({\rm Fig4d})_{nplanar} =- 2 g^2\! \int \frac{d^3 k}{(2\pi)^3}\frac{-k\cdot (k+p-r)+
  m^2}{[(k+p-r)^2+ M^2](k^2+M^2)} \cos[ 2k\wedge (p-r) + p\wedge r-
  q\wedge s], 
\end{equation}

\noindent
from which we obtain the following divergent part as $p \to r$:
\begin{equation}
\mbox{Divergent part of $({\rm Fig4d})_{nplanar}$} =-\frac{i g^2}{2\pi |\tilde p -\tilde r|}
\end{equation}

\noindent
so that, to cancel the divergence in ${(\rm Fig4c)}_{nplanar}$, we must set $g=e^2$.

We can check that all one-loop  additional diagrams containing
the vertex (\ref{4}) do not generate nonintegrable singularities.
Therefore, we may conclude that the model whose action is
\begin{equation}
S_{gauge}+ S_{scalar}+ S_{fermion} + S_{Yukawa}
\end{equation}

\noindent
is free from dangerous infrared divergences if $g= e^2$.

\section{Adjoint representation}
\label{section4}
\begin{figure}
\begin{center}\includegraphics{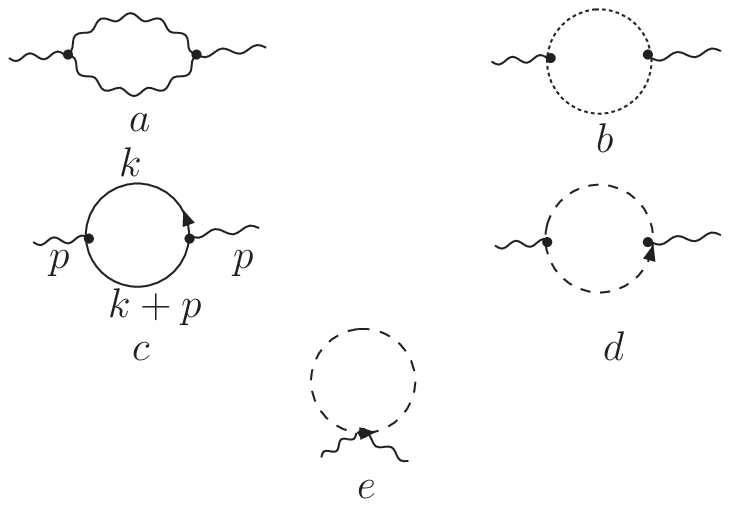}
\end{center}
\caption{One-loop corrections to the gauge field two point vertex function.}
\label{Fig2}
\end{figure}

\begin{figure}
\begin{center}\includegraphics{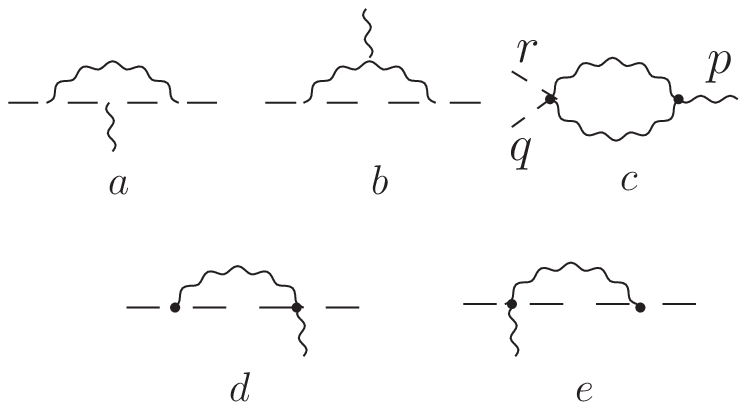}
\end{center}
\caption{One-loop corrections to the gauge-scalar field three point functions.}
\label{Fig3}
\end{figure}

\begin{figure}
\begin{center}\includegraphics{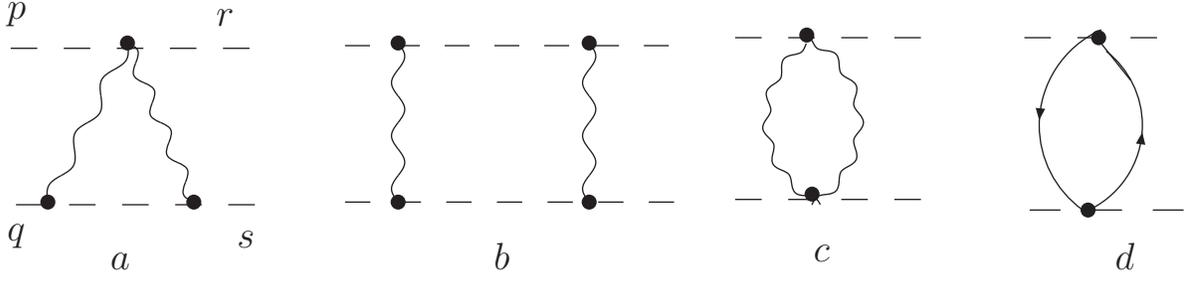}
\end{center}
\caption{One-loop contributions to the four point function.}
\label{Fig4}
\end{figure}

\begin{figure}
\begin{center}\includegraphics{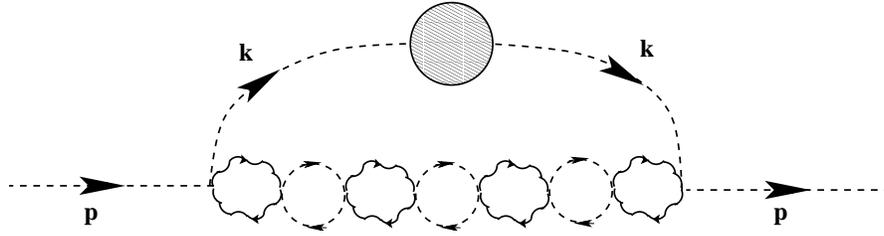}
\end{center}
\caption{Nonintegrable singularity generated by iteration of the graph
in Fig. \ref{Fig4}$c$.}
\label{Fig5}
\end{figure}

Let us now examine the  models introduced in the previous section but
with
the matter fields in the adjoint representation. We begin the
analysis by considering the model with action $S_{gauge}+ S_{fermion}$. 
In this case the graphs contributing to the two point proper vertex
functions
are no longer purely planar. Actually we have:

1. Gauge field two point proper vertex function. The relevant diagram
   is the graph in Fig.~\ref{Fig2}$c$ which yields
\begin{eqnarray}
\pi^{\mu \nu }_f(p)&=&-4 e^2\int \frac{d^{3}k}{(2\pi )^{3}}{\textrm{Tr}}
[\gamma ^{\nu }\frac{i}{-i\not \! k+M}\gamma ^{\mu }\frac{i}{-i(\not \! k+
\not \! p)+M}]\sin ^{2}{(k\wedge p)}\nonumber\\[16pt]
&=& \pi^{\mu \nu }_{f,\,planar}(p)+\pi^{\mu \nu }_{f,\,nplanar}(p)\label{NCfermionloop},
\end{eqnarray}
 
\noindent
where the subscript $f$ designates the fermionic contribution and the
planar and nonplanar parts are ($a = \sqrt{M^2 + x(1-x)p^2}$)

\begin{equation}
\pi^{\mu\nu}_{f,planar}=-\frac{ie^2}{\pi }(g^{\mu \nu }p^{2}-p^{\mu }p^{\nu })\int _{0}^{1}dx
\frac{x(1-x)}{a}+\frac{Me^2}{2\pi }\epsilon ^{\mu \nu \rho }p_{\rho }
\int _{0}^{1}dx\frac{1}{a}\label{left}
\end{equation}

\noindent
and

\begin{eqnarray}
&&\!\!\!\!\!\!\!\!\!\!\!\!\!\!\!\!\!\!\!\!\!\!\!\!\!\!\!\!\!\!\pi^{\mu \nu }_{f,\,nplanar}(p)  =  \frac{ie^2}{\pi }(g^{\mu \nu }p^{2}
-p^{\mu }p^{\nu })
\int _{0}^{1}dx\frac{x(1-x)}{a}{\rm e}^{-a\sqrt{\tilde{p}^{2}}}
\nonumber \\
 &  &+\frac{ie^2}{\pi }\frac{\tilde{p}^{\mu }\tilde{p}^{\nu }}{\tilde{p}^{2}}
\int _{0}^{1}dx(a+\frac{1}{\sqrt{\tilde{p}^{2}}}){\rm e}^{-a
\sqrt{\tilde{p}^{2}}} -\frac{Me^2}{2\pi }\epsilon ^{\mu \nu \rho }p_{\rho }\int _{0}^{1}dx
\frac{1}{a}{\rm e}^{-a\sqrt{\tilde{p}^{2}}},\label{n11b}
\end{eqnarray}

\noindent
which diverges linearly as $p \rightarrow 0$. To cancel this
divergence we add scalar fields described by the action in
Eq. (\ref{2}) but with mass $m=M$. We then have the contributions from the graphs in
Figs. \ref{Fig2}$d$ and \ref{Fig2}$e$ which  give
\begin{eqnarray}
\pi_{b}^{\mu \nu }(p)=-\frac{ie^2}{4\pi } &  &  \left\{ (g^{\mu \nu }
p^2-
p^{\mu }p^{\nu }) \int _{0}^{1}dx\frac{(1-2x)^2}{a}(1-{\textrm{e}}^{-a\sqrt{\tilde{p}^{2}}})
\right.\nonumber \\
 &  & \left.+4\frac{\tilde{p}^{\mu }\tilde{p}^{\nu }}{\tilde{p}^{2}}
\int _{0}^{1}dx(\frac{1}{\sqrt{\tilde{p}^{2}}}+a){\textrm{e}}^{-a
\sqrt{\tilde{p}^{2}}}\right\} .\label{n11a}
\end{eqnarray}
\noindent 
As we see, this last expression presents
the same infrared divergence as in the fermion case. Thus, as the masses
are equal the two divergences cancel.

\begin{figure}
\begin{center}\includegraphics{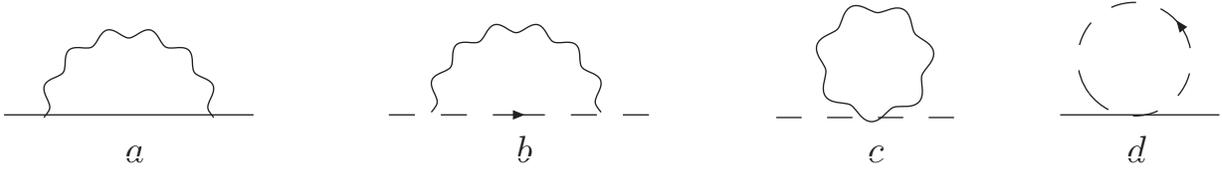}
\end{center}
\caption{One-loop corrections to the matter fields two point functions}
\label{Fig6}
\end{figure}

Let us now consider the one-loop corrections to the two point vertex
functions of the matter fields. Up to this point, the relevant
diagrams are depicted in Fig. \ref{Fig6}($a-c$).  In the Landau gauge
the integrands for  the
diagrams \ref{Fig6}$b$ and \ref{Fig6}$c$ vanish, so that the two point
vertex function of the scalar field does not introduce nonintegrable
infrared
singularities. Concerning the two point vertex function of the
fermion field,  after a
straightforward simplification, the graph in \ref{Fig6}$a$ furnishes

\begin{equation}
\Sigma_{\psi}=4ie^{2}\int \frac{d^{3}k}{(2\pi )^{3}}\frac{\not k
\left[i(k-p)_{\beta }\gamma ^{\beta }+m\right]}{k^{2}[(k-p)^{2}+m^{2})]}[1-\cos (2k\wedge p)],
\end{equation}

\noindent
whose nonplanar part yields

\begin{equation}
\Sigma_{\psi\,\, nplanar}= 
4e^{2}\int \frac{d^{3}k}{(2\pi )^{3}}\frac{\cos
(2k\wedge p) }{(k^{2}+m^{2})}+\mbox{finite term}=-\frac{ie^2}{\pi \sqrt{\tilde p^2}} {\rm
  e}^{-m\sqrt{\tilde p^2}} +\mbox{finite term}.
\end{equation}

\noindent
To keep things in perspective, we should recall that, besides this divergence we need also to cancel the one associated with
the four point function $<T\varphi^\dag \varphi\varphi^\dag \varphi>$. Before adding fermions this function receives contribution from the diagram
in the Fig.  \ref{Fig4}$c$. In the adjoint representation this
graph   gives
\begin{equation}
{\rm Fig4c}= -8 e^4\int
  \frac{d^3k}{(2\pi)^3}\frac{\epsilon_{\mu\rho\nu} 
k^\rho}{k^2}\frac{\epsilon^{\nu\alpha\mu} 
(k+p-r)_\alpha}{(k+p-r)^2} {\cal C}=-16 e^4
\int
  \frac{d^3k}{(2\pi)^3}\frac{(k+p-r)\cdot k}{(k+p-r)^2 k^2} {\cal C}\label{6}
\end{equation}

\noindent
where ${\cal C}$ is the trigonometric factor
\begin{eqnarray}
{\cal C} &=&[\sin(k\wedge q+ s\wedge q)\sin(k\wedge s)+\sin (k\wedge
  q)\sin(k\wedge s+s\wedge q)]\nonumber\\
&& \times [\sin(k\wedge r+ p\wedge r)\sin(k\wedge p)+\sin (k\wedge
  r)\sin(k\wedge p+p\wedge r)].\label{5}
\end{eqnarray}

\noindent
As done in our study of the fundamental representation,  we 
investigate the possibility to cancel these divergences by adding a
Yukawa like interaction. The structure of the trigonometric factor in
Eq. (\ref{5})
suggests that one should include the interaction

\begin{equation}
S_{Yukawa,\,adjoint}= g_1 \int d^3 x \{[\varphi^\dagger,\,
  \psi\,]_{*} \ast [\varphi,\, \overline \psi]_{*}- [\varphi^\dagger,\, \overline
  \psi\,]_{*} \ast [\varphi,\, \psi]_{*}\} .
  \end{equation}

\noindent
In fact, this interaction introduces a new vertex which will be still
represented by the last vertex in Fig. \ref{Fig1} but which 
corresponds to
\begin{equation}
\Gamma_{1\varphi\psi} = 4i g_1 [\sin(k_1\wedge p_1) \sin(k_2\wedge
  p_2)+
\sin(k_1\wedge p_2) \sin(k_2\wedge p_1)]. 
\end{equation}
Because of this new vertex, there is one additional diagram,
Fig. \ref{Fig6}$d$,  which provides the following
 contribution to the two point vertex function of the fermion field
\begin{equation}
{\rm Fig6d}= 8 g_1\int \frac{d^3k}{(2\pi)^3} \frac{\sin^2(k \wedge
  p)}{k^2 + m^2}.
 \end{equation}

As the nonplanar part of this graph is equal to
\begin{equation}
({\rm Fig6d})_{nplanar} =- 4 g_1 \int \frac{d^{3}k}{(2\pi )^{3}}\frac{\cos
(2k\wedge p) }{(k^{2}+m^{2})}=\frac{ig_1}{\pi \sqrt{\tilde p^2}} {\rm
  e}^{-m\sqrt{\tilde p^2}},
\end{equation}

\noindent
we see that the infrared singularity will cancel if $g_1=  e^2$.

Concerning the four point proper function of the scalar field, notice that  there is also a new diagram
which topologically is the same as the graph in Fig. \ref{Fig4}$d$ but
whose analytical expression is

\begin{eqnarray}
{\rm Fig4d}&=& 16 g_{1}^{2} \int \frac{d^{3}k}{(2\pi )^{3}}
{\rm Tr}[\Delta_\psi(k)\Delta_\psi(k+p-r)]{\cal C}\nonumber \\
&=&
-32g_{1}^{2} \int \frac{d^{3}k}{(2\pi )^{3}}\frac{-k\cdot (k+p-r) +
  m^2}{[(k+p-r)^2+m^2][k^2+m^2]}{\cal C}.\label{7}
\end{eqnarray}
\noindent
As $g_1= e^2$ the two contributions, Eqs. (\ref{6}) and (\ref{7}), do not cancel and
a linear IR divergence persists. To remove such divergence a further
extension of the model is needed. Taking into account these
observations, in the next section we will consider a superfield
CS model.

\section{the superfield CS model}
\label{section5}
We begin our analysis by considering the 2+1 dimensional superfield
CS model which is defined by the action 
\cite{SGRS}
\begin{eqnarray}
\label{chs}
S=m\int d^5 z (A^{\alpha}
*W_\alpha+\frac{i}{6}\{A^{\alpha},A^{\beta}\}_\ast\ast D_{\beta}A_{\alpha}+\frac{1}{12}
\{A^{\alpha},A^{\beta}\}_\ast*\{A_{\alpha},A_{\beta}\}_\ast)\,, \label{2n}
\end{eqnarray}
 where
\begin{eqnarray}
\label{sstr}
W_\beta =\frac{1}{2}D^\alpha D_\beta A_\alpha -
\frac{i}{2}[A^\alpha ,D_\alpha A_\beta ]_\ast-
\frac{1}{6}
[A^\alpha ,\{A_\alpha ,A_\beta \}_\ast]_\ast
\end{eqnarray}
is a superfield strength constructed from the spinor
superpotential $A_\alpha $. This action is invariant under the infinitesimal gauge transformations
\begin{equation}
\label{gt}
\delta A_{\alpha}=D_{\alpha}K-i[A_{\alpha},K]_\ast\,,
\end{equation}

\noindent
where $K$ is a scalar  superfield parameter. As a first step for
quantization, we eliminate this gauge freedom by choosing the 
gauge fixing and associate Faddeev-Popov terms as specified by the  action
\begin{eqnarray}
S_{GF+FP}=-\frac{m}{2\xi}\int d^5 z (D^\alpha A_\alpha )
(D^\beta A_\beta )+\frac{1}{2g^2}\int d^5 z (c'D^\alpha D_\alpha c+ic'*
D^\alpha [A_\alpha ,c]_\ast)\,,
\end{eqnarray}
so that the  quadratic part  of the action reads
\begin{eqnarray}
\label{s2a}
S_2=-\frac{1}{2}m\int d^5 z
A_\beta\Big[D^{\alpha}D^{\beta}+\frac{1}{\xi}D^{\beta}D^{\alpha}
\Big]A_\alpha +\frac{1}{2g^2}\int d^5 z c'D^\alpha D_\alpha c\,.
\end{eqnarray}

\noindent
From this action we get the free gauge and ghost propagators as being
\begin{eqnarray}
\label{pr1}
<A^\alpha (z_1)A^\beta (z_2)>=\frac{i}{4m\Box}\left[
D^{\beta}D^{\alpha}+\xi D^{\alpha}D^{\beta}
\right]\delta^5(z_1-z_2)\,,
\end{eqnarray}

\noindent
and
\begin{eqnarray}
\label{pr2}
<c'(z_1)c(z_2)>=-ig^2\frac{D^2}{\Box}\delta^5(z_1-z_2)\,.
\end{eqnarray}

The interaction part of the action determines three types of vertices:
\begin{eqnarray}
\label{val}
\Gamma_3&=&a_3 m A^{\beta}(k_1)A^{\alpha}(k_2)D_{\alpha}A_{\beta}(k_3)\sin(k_2\wedge k_3),
\nonumber\\
\Gamma_4&=&a_4 m A^{\beta}(k_1)
A^{\alpha}(k_2)A_{\alpha}(k_3)A_{\beta}(k_4)\sin(k_1\wedge k_2)
\sin(k_3\wedge k_4),\nonumber\\
\Gamma_c&=&-\frac{1}{g^2}c'(k_1)D^{\alpha}(A_{\alpha}(k_2)c(k_3))
\sin(k_2\wedge k_3),
\end{eqnarray}

\noindent
where $a_3=\frac23$ and $a_4=\frac13$. Instead of writing their
explicit values, we will retain the notations $a_3$ and $a_4$
to keep track of  the contributions of each vertex.

To study the divergence structure of the model we shall start by
determining the superficial degree of divergence $d(\gamma)$ associated to
a generic supergraph $\gamma$. Explicitly, $d(\gamma)$ receives contributions from the
propagators and, implicitly, from the supercovariant derivatives.
This last dependence can be unveiled by the use of the conversion rule
\begin{eqnarray}
D_{\alpha}(-k, \theta)D_{\beta}(-k, \theta)=k_{\alpha\beta}-C_{\alpha\beta}D^2(-k, \theta)\label{idsg}
\end{eqnarray}

\noindent
and the identity $(D^2)^2=-k^2$. Let $V_1$ be the number of
pure gauge vertices containing one super-derivative and $V_c$ the number of
ghost vertices; let $P_A$ and $P_c$ be the numbers of gauge and ghost
superpropagators and let $N_D$ be the number of supercovariant
derivatives that act on the external lines after the usual D-algebra
transformations. The superficial degree of divergence is then
\begin{eqnarray}
\label{o1}
d(\gamma)=2L+\frac{1}{2}(V_1+V_c)-P_A-P_c-\frac{1}{2} N_D\,,
\end{eqnarray}

\noindent
where $L$ is the number of loops. As we are going to consider Green
functions of the gauge superfield only, then $V_c=P_c$. Using this 
and the topological identity relating the number of  lines,
the number of vertices and the number of loops, the
above formula can be rewritten as
\begin{eqnarray}
\label{o}
d(\gamma)=2-\frac{1}{2} E_A-\frac{1}{2} N_D,
\end{eqnarray}

\noindent
where $E_A$ denotes the number of external $A$ lines.

At one loop, due to symmetric integration, the superficially
logarithmically divergent contributions are actually finite. We have therefore to
examine only graphs that are potentially linearly divergent. They
contribute to the two point gauge superfield vertex function and are
depicted in  Fig. 7. First notice that the ghost contribution in
Fig. 7$c$ is the same as in noncommutative super-QED$_3$ so that we just quote
the result from \cite{ours}
\begin{equation}
\Gamma_{2c}=-\frac{1}{2}\int \frac{d^3p}{(2\pi)^3}d^2 \theta_1
\int\frac{d^3k}{(2\pi)^3}\frac{\sin^2(k\wedge p)}{k^2}
A^\beta (-p,\theta_1)A_\beta (p,\theta_1)\,+\,\cdots\,,\label{2c}
\end{equation}

\noindent
where the ellipsis stands for finite terms. The second contribution,
which comes from the tadpole graph in Fig. 7$b$, is also easily evaluated
giving
\begin{equation}
\Gamma_{2b}=\frac{3}{2}a_4(1-\xi)\int \frac{d^3p}{(2\pi)^3}d^2 \theta_1
\int\frac{d^3k}{(2\pi)^3}\frac{\sin^2(k\wedge p)}{k^2}
A^\beta (-p,\theta_1)A_\beta (p,\theta_1).\label{2b}
\end{equation}

\begin{figure}[ht]
\includegraphics{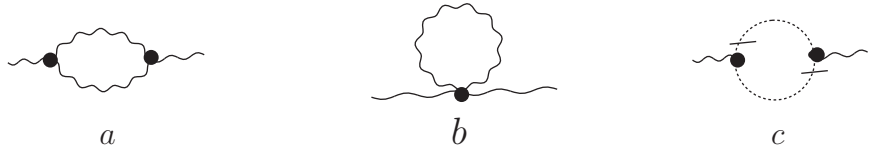}
\caption{Superficially linearly divergent diagrams contributing to the
two-point function of the gauge superfield.}
\label{Fig7}
\end{figure}

The evaluation of the graph in Fig. \ref{Fig7}$a$ is more complicated as it involves
two types of contractions distinguished by the fact that the two derivatives at
the vertices act on the same line (denoted by $(a)$, $(b)$ and $(c)$)
or on different lines (indicated by $(a^\prime)$, $(b^\prime)$ and
$(c^\prime)$):
\begin{eqnarray}
\Gamma_{2a} =(a) + (b)+ (c) + (a') + (b') + (c') 
\end{eqnarray}

\noindent
where
\begin{eqnarray}
\label{two1}
(a)&=&{m^2}a^2_3\int \frac{d^3p}{(2\pi)^3}d^2 \theta_1 d^2\theta_2
\int\frac{d^3k}{(2\pi)^3}\sin^2(k\wedge p)<D_{\alpha} A_{\beta}(k,\theta_1)D_{\alpha^\prime}A_{\beta^\prime}(-k,\theta_2)>\nonumber\\&\times&
<A^{\beta}(p-k,\theta_1)A^{\alpha^\prime}(-(p-k),\theta_2)>A^{\alpha}(-p,\theta_1)A^{\beta^\prime}(p,\theta_2),
\nonumber\\
(b)&=&\frac{m^2}{2}a^2_3\int \frac{d^3p}{(2\pi)^3}d^2 \theta_1 d^2\theta_2
\int\frac{d^3k}{(2\pi)^3}\sin^2(k\wedge p)
<D_{\alpha} A_{\beta}(k,\theta_1)D_{\alpha^\prime}A_{\beta^\prime}(-k,\theta_2)>\nonumber\\&\times&
<A^{\beta}(p-k,\theta_1)A^{\beta^\prime}(-(p-k),\theta_2)>A^{\alpha} (-p,\theta_1)A^{\alpha^\prime}(p,\theta_2),
\nonumber\\
(c)&=&\frac{m^2}{2}a^2_3\int \frac{d^3p}{(2\pi)^3}d^2 \theta_1 d^2\theta_2
\int\frac{d^3k}{(2\pi)^3}\sin^2(k\wedge p)
<D_{\alpha} A_{\beta}(k,\theta_1)D_{\alpha^\prime}A_{\beta^\prime}(-k,\theta_2)>\nonumber\\&\times&
<A^{\alpha} (p-k,\theta_1)A^{\alpha^\prime}(-(p-k),\theta_2)>A^{\beta}(-p,\theta_1)A^{\beta^\prime}(p,\theta_2),
\end{eqnarray}
\begin{eqnarray}
\label{two2}
(a')&=&m^2a^2_3\int \frac{d^3p}{(2\pi)^3}d^2 \theta_1 d^2\theta_2
\int\frac{d^3k}{(2\pi)^3}\sin^2(k\wedge p)
<D_{\alpha} A_{\beta}(k,\theta_1)A^{\alpha^\prime}(-k,\theta_2)>\nonumber\\&\times&
<A^{\beta}(p-k,\theta_1)D_{\alpha^\prime}A_{\beta^\prime}(-(p-k),\theta_2)>
A^{\alpha} (-p,\theta_1)A^{\beta^\prime}(p,\theta_2),\nonumber\\
(b')&=&\frac{m^2}{2}a^2_3\int \frac{d^3p}{(2\pi)^3}d^2 \theta_1 d^2\theta_2
\int\frac{d^3k}{(2\pi)^3}\sin^2(k\wedge p)<D_{\alpha} A_{\beta}
(k,\theta_1)A^{\beta^\prime}(-k,\theta_2)>
\nonumber\\&\times&
<A^{\beta}(p-k,\theta_1)D_{\alpha^\prime}A_{\beta^\prime}(-(p-k),\theta_2)>
A^{\alpha}(-p,\theta_1)A^{\alpha^\prime}(p,\theta_2),\nonumber\\
(c')&=&\frac{m^2}{2}a^2_3\int \frac{d^3p}{(2\pi)^3}d^2 \theta_1 d^2\theta_2
\int\frac{d^3k}{(2\pi)^3}\sin^2(k\wedge p)<D_{\alpha} A_{\beta}(k,\theta_1)A^{\alpha^\prime}(-k,\theta_2)>\nonumber\\&\times&
<A^{\alpha} (p-k,\theta_1)D_{\alpha^\prime}A_{\beta^\prime}(-(p-k),\theta_2)>
A^{\beta}(-p,\theta_1)A^{\beta^\prime}(p,\theta_2).
\end{eqnarray}
After straightforward D-algebra transformations we obtain
\begin{eqnarray}
(a)&=&8 a^2_3\xi\int d^2\theta\int\frac{d^3pd^3k}{(2\pi)^6}J k^2 
A^\beta (-p,\theta)A_\beta (p,\theta),\nonumber\\
(b)&=&8 a^2_3\xi\int d^2\theta\int\frac{d^3pd^3k}{(2\pi)^6}J k^2 
A^\beta (-p,\theta)A_\beta (p,\theta),\nonumber\\
(c)&=&4 a^2_3(\xi-\xi^2)\int d^2\theta\int\frac{d^3pd^3k}{(2\pi)^6}J k^2 
A^\beta (-p,\theta)A_\beta (p,\theta),\nonumber\\
(a')&=&8 a^2_3\xi \int d^2\theta\int\frac{d^3pd^3k}{(2\pi)^6}J k^2 
A^\beta (-p,\theta)A_\beta (p,\theta),\nonumber\\
(b')&=&8 a^2_3\xi\int d^2\theta\int\frac{d^3pd^3k}{(2\pi)^6}J k^2 
A^\beta (-p,\theta)A_\beta (p,\theta),\nonumber\\
(c')&=&4 a^2_3\xi^2\int d^2\theta\int\frac{d^3pd^3k}{(2\pi)^6}J k^2 
A^\beta (-p,\theta)A_\beta (p,\theta),
\end{eqnarray}
where 
\begin{eqnarray}
J=\frac{1}{32}\frac{\sin^2(k\wedge p)}{k^2(p-k)^2}.
\end{eqnarray}
The final contribution of this graph is therefore
\begin{equation}
\label{2a}
\Gamma_{2a}=\frac{9}{8}a^2_3\xi\int \frac{d^3p}{(2\pi)^3}d^2 \theta
\int\frac{d^3k}{(2\pi)^3}\frac{\sin^2(k\wedge p)}{k^2}
A^\beta (-p,\theta)A_\beta (p,\theta).
\end{equation}

Thus, collecting the results in (\ref{2c}), (\ref{2b}) and (\ref{2a})
we get that the would be divergent part of $\Gamma_2$,
\begin{eqnarray}
\Gamma_{2}^{Div}=(\frac{9}{8}a^2_3\xi+\frac{3}{2}a_4(1-\xi)-\frac{1}{2})
\int \frac{d^3p}{(2\pi)^3}d^2 \theta
\int\frac{d^3k}{(2\pi)^3}
\frac{\sin^2(k\wedge p)}{k^2}
A^\beta (-p,\theta)A_\beta (p,\theta),
\end{eqnarray}
vanishes irrespectively of the gauge parameter $\xi$. This means that
the one-loop two point vertex function of the gauge superfield is free
from both UV and UV/IR infrared singularities in any covariant gauge.
As a matter of fact, using   arguments similar to those presented in \cite{ours} one can demonstrate 
that all superficially logarithmically divergent graphs are finite. We
therefore conclude that {\em in any gauge} the model is one-loop finite.

Let us now consider the effect of the inclusion of matter fields. We
first examine the case in which a scalar superfield in the adjoint
representation couples to the CS superfield through
the action
\begin{eqnarray}
\label{matter}
S^A&=&\int d^5 z
\Big\{\bar\phi(D^2-M)\phi
-\frac{i}{2} (g[\bar\phi,A^{\alpha}]_**D_{\alpha}\phi-
g D^{\alpha}\bar\phi*[A_{\alpha},\phi]_*)
\nonumber\\&\phantom a &
-\frac{g^2}{2} [\bar\phi,A^{\alpha}]_**[A_{\alpha},\phi]_*
\Big\}.
\end{eqnarray}

\noindent
With this modification the  superficial degree of divergence in
Eq. (\ref{o}) must be replaced by
 \begin{eqnarray}
\label{o2}
d(\gamma)=2-\frac{1}{2}(E_A+E_{\phi})-\frac{N_D}{2},
\end{eqnarray}
where $E_A$ and $E_{\phi}$ are the numbers of the external $A$ and $\phi$
lines, respectively. The more dangerous situations correspond to 
 linearly divergent contributions which are possible only if 
$E_{\phi}=2$ or $E_A=2$.
The addition of the action (\ref{matter}) generates new contributions
to the two point proper vertex function of the gauge superfield.  The
corresponding supergraphs are listed in Fig. 8 and the details of
their computation are the same as in the three-dimensional
noncommutative $CP^{N-1}$ model \cite{cpn}.
They give the following contributions to the effective action

\begin{eqnarray}
\label{s1}
iS^A_{8a}(p)&=&-2g^2 \int d^2\theta \int \frac{d^3k}{(2\pi)^3} 
I(k,p)
\nonumber\\&\times&\Big[
(k^2+M^2)C_{\alpha\beta} A^{\alpha}(-p,\theta)A^{\beta}(p,\theta)
+(k_{\alpha\beta}+MC_{\alpha\beta})(D^2A^{\alpha}(-p,\theta)) A^{\beta}(p,\theta)
\nonumber\\&+&\frac{1}{2} D^{\gamma}D^{\alpha}A_{\alpha}
(-p,\theta)(k_{\gamma\beta}+MC_{\gamma\beta})A^{\beta}(p,\theta)
\Big]
\end{eqnarray}
and
\begin{eqnarray}
\label{s2}
iS^A_{8b}(p)&=&2g^2\int\frac{d^3k}{(2\pi)^3}\frac{\sin^2(k\wedge p)}{(k+p)^2+M^2}
C_{\alpha\beta} A^{\alpha}(-p,\theta)A^{\beta}(p,\theta),
\end{eqnarray}
where
\begin{equation}
I(k,p)=\frac{\sin^2(k\wedge p)}{(k^2+M^2)[(k+p)^2+M^2]}.
\end{equation}

Although individually divergent the sum of $iS^A_{8a}(p)$ and
$iS^A_{8b}(p)$ is finite being equal to

\begin{eqnarray}
\label{stot}
iS^A_8(p)&=&-2g^2 
\int d^2\theta \int \frac{d^3k}{(2\pi)^3} 
I(k,p)
\nonumber\\&\times&
(k_{\gamma\beta}+MC_{\gamma\beta})\Big[(D^2A^{\gamma}(-p,\theta)) A^{\beta}(p,\theta)
+\frac{1}{2} D^{\gamma}D^{\alpha}A_{\alpha}(-p,\theta) A^{\beta}(p,\theta)
\Big],
\end{eqnarray}

\noindent
or equivalently,
\begin{eqnarray}
\label{t2p1}
S^A_8(p)&=& \frac{g^2}{16\pi} \int d^2\theta f(p) A^{\beta}(p,\theta) [D^2+2M]
W_{0\beta}(-p,\theta)\nonumber\\& =& 
\frac{g^2}{16\pi} \int d^2\theta f(p)[ W_{0}^{\alpha} 
W_{0\alpha}+ 2M
W^{\alpha}_{0} A_\alpha], 
\end{eqnarray}
where
\begin{eqnarray}
f(p) = -16 \pi i \int \frac{d^3 k}{(2\pi)^3}I(k,p)
\end{eqnarray}
and $W^{\alpha}_{0}= \frac12 D^\beta D^\alpha A_\beta$ is a linearized superfield strength. As we see, these
graphs originate nonlocal Maxwell and CS terms in the
effective action.

Let us now consider the two point function of the scalar
superfield. The one-loop contributing graphs are depicted in Fig. 9;
they are superficially linearly divergent. Notice that, as before by
reasons of symmetry, the would be logarithmic divergences vanish
and therefore all terms which do not contain linear divergences are
finite.

The UV leading part of the graph in Fig. 9$a$, which involves two
vertices with three fields is
\begin{eqnarray}
iS^{(1)A}_{\phi\bar{\phi}}&=&-\frac{1}{2}g^2
\int\frac{d^3k}{(2\pi)^3}\int d^2\theta_1
d^2\theta_2\frac{\sin^2(k\wedge p)}{4mk^2[(p+k)^2+M^2]}
(D^{\beta}D^{\alpha}+\xi D^{\alpha}D^{\beta})\delta_{12}
\nonumber\\&\times&
D_{\alpha 1}(D^2+M)D_{\beta 2}\delta_{12}
[\phi(-p,\theta_1)\bar\phi(p,\theta_2)+\bar\phi(-p,\theta_1)
\phi(p,\theta_2)]\label{scalar}
\end{eqnarray}
and, after D-algebra transformations turns out to be 
\begin{eqnarray}
iS^{(1)A}_{\phi\bar{\phi}}=\xi g^2\int d^2\theta
\frac{\phi(-p,\theta)\bar\phi(p,\theta)}{m}\int\frac{d^3k}{(2\pi)^3}
\frac{\sin^2(k\wedge p)}{k^2}+\mbox{finite term}.\label{landau}
\end{eqnarray}
Notice that this gauge dependent contribution only vanishes in the
Landau, $\xi=0$ gauge, as could be anticipated from a rapid inspection
of Eq. (\ref{scalar}).  Now, after  trivial D-algebra
transformations the contribution from the graph in Fig. 9$b$ becomes
\begin{eqnarray}
iS^{(2)A}_{\phi\bar{\phi}}=-(1-\xi)g^2\int d^2\theta
\frac{\phi(-p,\theta)\bar\phi(p,\theta)}{m}\int\frac{d^3k}{(2\pi)^3}
\frac{\sin^2(k\wedge p)}{k^2}+\mbox{finite term}.\label{feynman}
\end{eqnarray}
Differently from Eq. (\ref{landau}) the above result only vanishes in
the Feynman, $\xi=1$, gauge where the propagator of the $A^\alpha$
superfield does not contain spinor derivatives. The sum of
Eqs. (\ref{landau}) and (\ref{feynman}) only vanishes in the $\xi=1/2$
gauge and thus only in this gauge the model with the matter superfields
 in the adjoint representation is free from dangerous  UV/IR  infrared
 divergences. 

A more favorable situation occurs if the
 matter superfield belongs to the fundamental representation of the
gauge group. In this case the matter action is
\begin{eqnarray}
\label{matter1}
S^F&=&\int d^5 z
\Big[\bar\phi(D^2-M)\phi
-\frac{ig}{2} (\bar\phi*A^{\alpha}*D_{\alpha}\phi-
D^{\alpha}\bar\phi*A_{\alpha}*\phi)
\nonumber\\&\phantom a &
-\frac{g^2}{2} \bar\phi*A^{\alpha}*A_{\alpha}*\phi
\Big],
\end{eqnarray}
which implies in the following form of the vertices after the Fourier 
transform:
\begin{eqnarray}
\Gamma^F_3&=&-\frac{ig}{2}A^{\alpha}(k_1)(D_{\alpha}\phi(k_2)\bar\phi(k_3)-
\phi(k_2)D_{\alpha}\bar{\phi}(k_3))e^{ik_2\wedge k_3},\nonumber\\
\Gamma^F_4&=&-\frac{g^2}{2}
\bar\phi(k_1)A^{\alpha}(k_2)A_{\alpha}(k_3)\phi(k_4)
e^{ik_1\wedge k_2+ik_3\wedge k_4}.\label{fund}
\end{eqnarray}
We can easily calculate the contributions of graphs containing these
vertices to the two point function of the gauge superfield. In fact,
the D-algebra transformations are exactly the same as in the adjoint 
representation, the only differences in the analytical expressions
being due to the replacement of trigonometric factors by phases in
the way specified in  the Eqs. (\ref{fund}). However, these phase
factors do not interfere with the calculations since both graphs turn
out to be planar. Their corresponding analytical expressions are
\begin{eqnarray}
\label{s1f}
iS^F_{8a}(p)&=&-\frac{g^2}{2}\int d^2\theta \int \frac{d^3k}{(2\pi)^3} 
\frac{1}{(k^2+M^2)[(k+p)^2+M^2]}
\nonumber\\&\times&\Big[
(k^2+M^2)C_{\alpha\beta} A^{\alpha}(-p,\theta)A^{\beta}(p,\theta)
+(k_{\alpha\beta}+MC_{\alpha\beta})(D^2A^{\alpha}(-p,\theta)) A^{\beta}(p,\theta)
\nonumber\\&+&\frac{1}{2} D^{\gamma}D^{\alpha}A_{\alpha}
(-p,\theta)(k_{\gamma\beta}+MC_{\gamma\beta})A^{\beta}(p,\theta)
\Big]
\end{eqnarray}
and
\begin{eqnarray}
\label{s2f}
iS^F_{8b}(p)&=&\frac{g^2}{2}\int\frac{d^3k}{(2\pi)^3}\frac{1}{(k+p)^2+M^2}
C_{\alpha\beta} A^{\alpha}(-p,\theta)A^{\beta}(p,\theta).
\end{eqnarray}
Their sum is also finite and equal to
\begin{eqnarray}
\label{stot1}
iS^F_8(p)&=&-\frac{g^2}{2} \int d^2\theta \int \frac{d^3k}{(2\pi)^3} 
\frac{1}{(k^2+M^2)[(k+p)^2+M^2]}
\nonumber\\&\times&
(k_{\gamma\beta}+MC_{\gamma\beta})\Big[(D^2A^{\gamma}(-p,\theta)) 
A^{\beta}(p,\theta)
+\frac{1}{2} D^{\gamma}D^{\alpha}A_{\alpha}(-p,\theta) A^{\beta}(p,\theta)
\Big],
\end{eqnarray}
the only difference with respect to Eq. (\ref{stot}) being the absence
of the trigonometric factor.

We still have to examine the contributions to the two point vertex
function of the scalar superfield. The relevant graphs are again those
drawn in Fig. \ref{Fig9} and in this case are totally planar. We get
\begin{eqnarray}
iS^{(1)F}_{\phi\bar{\phi}}&=&-\frac{1}{8}g^2
\int\frac{d^3k}{(2\pi)^3}\int d^2\theta_1
d^2\theta_2\frac{1}{4mk^2[(p+k)^2+M^2]}
(D^{\beta}D^{\alpha}+\xi D^{\alpha}D^{\beta})\delta_{12}
\nonumber\\&\times&
D_{\alpha 1}(D^2+M)D_{\beta 2}\delta_{12}
[\phi(-p,\theta_1)\bar\phi(p,\theta_2)+\bar\phi(-p,\theta_1)
\phi(p,\theta_2)],
\end{eqnarray}
which after D-algebra transformations becomes
\begin{eqnarray}
iS^{(1)F}_{\phi\bar{\phi}}=\frac{1}{4}g^2\xi\int d^2\theta
\frac{\phi(-p,\theta)\bar\phi(p,\theta)}{m}\int\frac{d^3k}{(2\pi)^3}
\frac{1}{k^2}+\mbox{finite term}.
\end{eqnarray}

The D-algebra transformations for the second graph are
simpler and yield
\begin{eqnarray}
iS^{(2)F}_{\phi\bar{\phi}}=-\frac{1}{4}(1-\xi)g^2\int d^2\theta
\frac{\phi(-p,\theta)\bar\phi(p,\theta)}{m}\int\frac{d^3k}{(2\pi)^3}
\frac{1}{k^2}.
\end{eqnarray}
In the context of dimensional regularization, which we are implicitly
assuming, these divergent parts vanish. Thus in any gauge the one-loop
contributions to the two point vertex function of the scalar
superfield are finite. This result singles out the fundamental
representation as the preferable one for the construction of the model.

It should be noticed that although absent in the one-loop corrections
a quartic self-interaction of the scalar superfield may be induced at
higher orders. In that situation for renormalizability one should a
fortiori introduce the coupling
\begin{eqnarray}
\label{add}
-\frac{\lambda}{2}\int d^5 z\bar\phi*\phi*\bar{\phi}*\phi,
\end{eqnarray}

\noindent
which in its turn generates new one-loop graphs. In particular, for
 the two point function of the scalar superfield we have the graph
 depicted in Fig. \ref{Fig10} which corresponds to

\begin{eqnarray}
-2\lambda\int d^2\theta\int\frac{d^3p}{(2\pi)^3}
\int\frac{d^3k}{(2\pi)^3}\frac{1}{k^2+M^2}(D^2+M)\delta_{11}
\phi(-p,\theta)\bar\phi(p,\theta),
\end{eqnarray}
which after a trivial D-algebra transformation is equal to
\begin{eqnarray}
-2\lambda\int d^2\theta\int\frac{d^3p}{(2\pi)^3}
\int\frac{d^3k}{(2\pi)^3}\frac{1}{k^2+M^2}
\phi(-p,\theta)\bar\phi(p,\theta),
\end{eqnarray}

\noindent
providing a finite mass renormalization for the scalar superfield.
\section{Conclusions}
\label{section6}
In this work we have studied various models of matter fields coupled to the
CS field both in the fundamental and in the
adjoint representation of the $U(1)$ noncommutative gauge
group. Special attention was given to the occurrence of UV/IR mixing
as it  may generate nonintegrable infrared singularities. We began by
proving that the model describing a fermionic field minimally coupled
to the CS field is free from dangerous UV/IR mixing. On the other hand,
the model with only  a scalar field also in the fundamental
representation and  minimally coupled to the CS field
presents a linear infrared divergence in the one-loop contribution to
the four point vertex function of the matter field. We proved that it
is
possible to cancel such divergence by incorporating fermions
interacting with the scalar field via a noncommutative Yukawa like
Lagrangian. The situations are more complicated if the
matter
fields belong to the adjoint representation: to eliminate the
UV/IR mixing in the one-loop contributions to the gauge field
propagator it is necessary to consider a more general model
containing both scalar and fermionic fields minimally coupled to the
CS field. However, even with the addition of a Yukawa interaction it
was not possible to eliminate all one-loop infrared divergences which
are present in the two point vertex function of the fermionic field
and
also in the four point function of the scalar field. More general
interactions
seemed to be necessary and also motivated by results in supersymmetric 
gauge theories \cite{ours,cpn} we were led to study a noncommutative
CS superfield coupled to matter.  We first
demonstrate
that the pure gauge sector is finite in an arbitrary gauge. 
The inclusion of matter  brought new features  depending on the
representation to which the corresponding superfield belongs. For
the matter superfield in the fundamental representation of the gauge
group all  one-loop graphs with positive superficial degree of divergence
are planar and are therefore finite in the context of
dimensional regularization. However, for the matter in the adjoint
representation we found that the absence of dangerous UV/IR singularities
in the two point vertex function of the matter field only happens in a
particular gauge, namely $\xi=1/2$.

\section{Acknowledgments} 
This work was partially supported by Funda\c{c}\~{a}o de Amparo 
\`{a} Pesquisa do Estado de S\~{a}o Paulo (FAPESP) and Conselho 
Nacional de Desenvolvimento Cient\'{\i}fico e Tecnol\'{o}gico (CNPq). 
A. Yu. P. has been supported by FAPESP, project No. 00/12671-7.

\newpage

\begin{figure}[ht]
\includegraphics{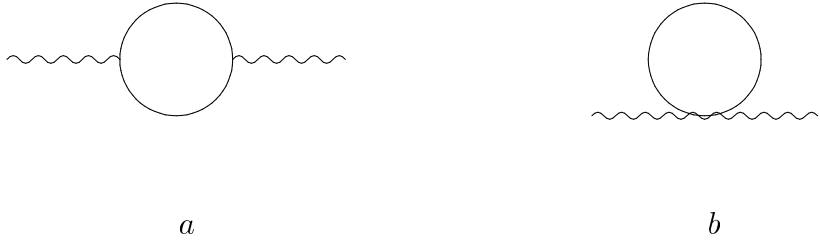}
\caption{Coupling to matter: contributions to the two-point function
of gauge superfield.}
\label{Fig8}
\end{figure}

\begin{figure}[ht]
\includegraphics{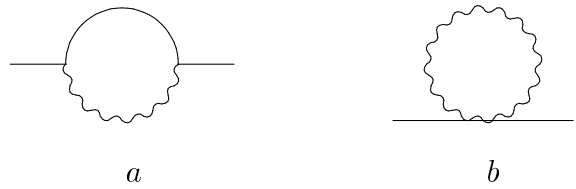}
\caption{Coupling to matter: contributions to the two-point function
of matter field.}
\label{Fig9}
\end{figure}

\begin{figure}[ht]
\includegraphics{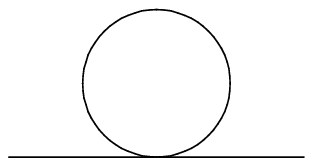}
\caption{The contribution to the two-point function
of matter field generated by the matter self-interaction.}
\label{Fig10}
\end{figure}


\begin{thebibliography}{10}
\bibitem{Kaminsky} K. Kaminsky, Nucl. Phys. {\bf B679}, 189 (2004);
  K. Kaminsky, Y. Okawa and H. Ooguri, Nucl. Phys. {\bf B663}, 33 (2003).
\bibitem{Martin} C. P. Martin, Phys. Lett. B {\bf 515}, 185 (2001).
\bibitem{Dayi} O. F. Dayi, Phys. Lett. B {\bf 560}, 239 (2003).

\bibitem{Chandra} B. Chandrasekhar and P. K. Panigrahi, J. High Energy Phys. 
03 (2003) 015. 
\bibitem{Anacleto} M. A. Anacleto, M. Gomes, A. J. da Silva,
D. Spehler, Phys. Rev D {\bf 70}, 085005 (2004);
L. C. T. Brito, M. Gomes, S. Perez, A. J. da Silva, 
J. Phys. {\bf A37}, 9989 (2004).

\bibitem{Saha} P. Mukherjee, A. Saha, ``A new approach to the analysis
of the noncommutative Chern-Simons theory'', hep-th/0409248.

\bibitem{Polychronakos} V. P. Nair and  A. P. Polychronakos,  
Phys. Rev. Lett. 87, 030403 (2001).

\bibitem{Jabbari} M. M. Sheikh-Jabbari, Phys. Lett. B {\bf 510}, 247 (2001).

\bibitem{Wu} G. H. Chen and Y. S. Wu, Nucl. Phys. {\bf B628}, 473 (2002).

\bibitem{Park} D. Bak, K. Lee and J. H. Park, Phys. Rev. Lett. 87,
  030402 (2001).

\bibitem{Chu} C. S. Chu, Nucl. Phys. {\bf B580}, 352 (2000).

\bibitem{Silva} N. Grandi and G. A. Silva, Phys. Lett. B {\bf 507}, 345
  (2001).

\bibitem{Das} A. Das and M. M. Sheikh-Jabbari, J. High Energy Phys. 
06 (2001) 028.

\bibitem{Schweda} A. A. Bichl, J. M. Grimstrup, V. Putz, M. Schweda, 
J. High Energy Phys.  07 (2000) 046.

\bibitem{Pani} P. K. Panigrahi, T. Shreecharan, ``Induced Magnetic
  Moment in Noncommutativa Chern-Simons Scalar QED'', hep-th/0411132.

\bibitem{Lee} C. Lee, K. Lee, E. Weinberg, Phys. Lett. B {\bf 243}, 105
(1990); H.-J. Kao, K. Lee, C. Lee, T. Lee, Phys. Lett. B {\bf 341}, 181 (1994).

\bibitem{Gates} S. J. Gates, H. Nishino, Phys. Lett. B {\bf 281}, 72 (1992).

\bibitem{Avdeev} L. V. Avdeev, G. V. Grigoryev, D. I. Kazakov,
Nucl. Phys. {\bf B382}, 561 (1992); L. V. Avdeev, D. I. Kazakov,
I. N. Kondrashuk, Nucl. Phys. {\bf B391}, 333 (1993).

\bibitem{Gomis} J. Gomis and T. Mehen, Nucl. Phys. {\bf B591}, 265 (2000); 
D. Bahns, S. Doplicher, K. Fredenhagen and G. Piacitelli,
Phys. Lett. B {\bf B533}, 178 (2002).

\bibitem{SGRS} S. J. Gates, M. T. Grisaru, M. Rocek,
W. Siegel. Superspace or 
One Thousand and One Lessons in Supersymmetry. Benjamin/Cummings, 1983.
\bibitem{ours} A. F. Ferrari, H. O. Girotti, M. Gomes, A. Yu. Petrov,
A. A. Ribeiro, A. J. da Silva, Phys. Lett. B {\bf 577}, 83 (2003).
\bibitem{cpn} E. A. Asano, H. O. Girotti, M. Gomes, A. Yu. Petrov,
A. G. Rodrigues, A. J. da Silva, Phys. Rev. D {\bf 69}, 105012 (2004).
\end{thebibliography}
\end{document}